\documentclass[12pt]{elsart}
\usepackage {epsfig}
\usepackage[english]{babel}
\usepackage[T1]{fontenc}
\usepackage[latin2]{inputenc}
\usepackage{times}
\begin{document}
\begin{frontmatter}
\title{Fundamental building blocks of eumelanins:
electronic properties of indolequinone-dimers.}
\author{K. Bochenek \and E. Gudowska--Nowak}
\address{Marian Smoluchowski Institute of Physics,\\ 
Jagellonian University, Kraków, Poland}
\begin{abstract}
We present results from the theoretical INDO calculations of the electronic
structure for stacked eumelanins' monomers. As basic indolic components of the eumelanin
structure 5,6-dihydroxyindole (DHI or HQ) and its oxidized forms (SQ and IQ) were chosen.
The results reveal dependency of electronic properties of such aggregates on monomers' 
redox states. They point out also a tendency to localize an extra charge on one of 
dimer's subunits that could be suggestive of an electron hopping  as a model mechanism for
the electron transfer in eumelanins.
\end{abstract}
\end{frontmatter}
\section{Introduction}

\begin{figure}
\begin{center}
\epsfig{file=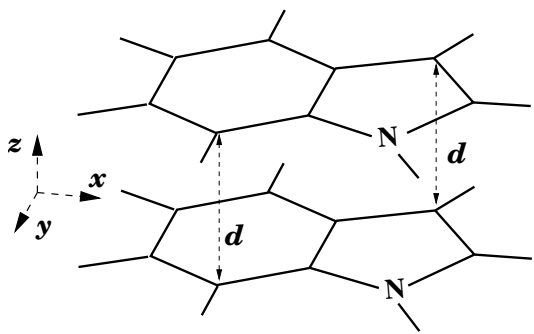,width=7cm}
\caption{An example of the examinated aggregate of the monomers.}
\label{monomers}
\end{center}
\end{figure}

Melanin is a photoactive biopolymer \cite{prota1}, a common biological pigment responsible
 for much of coloration observed in nature.
Despite its ubiquity, documented photoprotective \cite{sarna,zeise} character, reported photosensitizing
activity \cite{zhang,young}
 and otherwise hypothesized
multifunctionality of the pigment in animals and humans \cite{prota1,prota2},  neither the size nor
 the structure of the fundamental
molecular unit of the melanins has been well understood yet.
Black eumelanins are composed of indolic units derived from the oxidation of tyrosine: among those
5,6-dihydroxyindole (DHI or HQ) together with their oxidized forms
(indole-5,6-quinone 
(IQ)) and semi\-quinone (SQ) and 5,6-\-di\-hydroxy\-indole-2-carboxylic acid (DHICA) 
are usually accepted \cite{prota1,sarna,zhang,zajac,cheng} as main components of natural eumelanins
and their synthetic analogues. \\
Photoreactivity, ability to bind transition metal ions, redox properties and presence 
of persistent 
free radical centers in the pigment structure are usually mentioned as main features of eumelanins 
\cite{sarna}, although their biological significance is still unclear.\\
X-ray diffraction measurements of dried synthetic pigments have led to a model that pictures
eumelanin \cite{zajac,cheng,rosei,clancy} as a $\pi$ - stack of  crosslinked planar oligomeric 
structures with a spacing of $\approx 3.4$\AA. Such a $\pi$-conjugated, randomly arranged 
heteropolymer system could be responsible for intrinsic semiconductivity of eumelanins
\cite{sarna,prota2,pullman,galvao1}.\\
The eumelanins' monomers, as well as their dimers and polymers have been
  subject to theoretical
studies in the framework of H\"uckel theory \cite{pullman,galvao1,galvao2} that have supported
the hypothesis 
of eumelanins as semiconductors. A further investigation along those lines has been
 performed for the charged monomers \cite{marinez} where the stability of charged HQ, SQ 
and IQ was investigated. Following that study all monomers have been concluded to be strong electron 
acceptors (in the 
decreasing order from SQ to IQ to HQ). Elucidation of the spectral properties of synthetic eumelanins 
has been also addressed
\cite{marinez,my} pointing out  consequences of the assumed random organisation of the polymer that
explains well experimental absorption spectra of synthetic eumelanins.\\
 The aim of the present study is to examine electronic properties: orbital- and 
charge- localization  of stacked 
eumelanins' monomers forming an "indolic sandwich" with a characteristic separation
distance of about $3$\AA.  Two parallel, mostly planar 
monomers (Figure~\ref{monomers}) laying one above the other are chosen as a main unit for 
modelling.
A few assemblies in which one of the monomers has been rotated or moved apart
have been also examined checking for the effect of conformational variations on formation of a
stable dimeric unit.\\ 
Natural
eumelanins are built of both DHI and DHICA \cite{prota2} monomers but it is possible 
to prepare synthetic material as pure DHI-- or DHICA--melanin \cite{clancy}.
In this respect, the  article focuses on a distinct class of synthetic eumelanins discussing
properties of simple DHIs aggregates only.
\section{Methods}
Since experimental data on the geometry of the monomers are not available, the model structures 
have been created in the InsightII/MSI module \cite{insight} and geometrically optimized by use 
of the consistent valence forcefield (CVFF) \cite{insight}  until the demanded covergence in energy has been achieved (the maximum derivative
 less than 0.0001 kcal mol$^{-1}$ \AA$^{-1}$).
Coordinates of such monomers have been used as input data for semiempirical INDO 
calculations \cite{ridley}. Ground electronic states of the neutral charged molecules,
as well as of the negatively charged molecules from wich one H$^+$ ion was removed (the 
number of electrons remains the same),
have been obtained as closed-shell molecular orbital wave functions in the Hartree-Fock (RHF)
framework. For molecules with an extra electron or without one elctron Unrestricted 
Hartree-Fock (UHF) method has been used.\\
For molecules without hydrogen ion the first excited states were calculated 
by 
configuration interactions (CI)  among configurations generated as single excitations
from RHF ground state. The CI method included the highest 10(20) occupied and
 lowest 10(20) unoccupied molecular orbitals for the monomers(dimers). \footnote{Our test calculations have
 revealed that expansion of the CI basis to more than 20 highest occupied nad
 20 lowest unoccupied  orbitals does not change significantly neither the values
 of estimated excitation energies nor the main contributions $c_{ik}$ to the
 molecular orbitals of the dimer.} 
At the first step, the calculations have been performed for each monomer (HQ, IQ and SQ)
and then for the aggregates built of two stacked monomers (HH, II, SS, SH, HI, 
IS). For simplicity, we adopt here a notation where HH, II, SH, {\it etc.} stand
for dimers of HQ-HQ, IQ-IQ and SQ-HQ, respectively. The distance between 
molecules has been controlled within the range 2.86-4.12\AA{} by a translation (a
shift in the $z$-coordinate) of all atoms forming a displaced monomer in the
pair. 
In order to test whether the mirror symmetry of dimer coordinates in HH, II and SS assemblies
  does not generate
special results in the INDO runs,  
 model HQ-HQ aggregates  were 
built in which one monomer has been optimized with the CVFF and the second 
with a slightly different version of the consistent valence forcefiled
\cite{insight} 
intended
for application to polymers and organic materials (PCFF). 
Such a procedure breaks the symmetry of the aggregate producing nevertheless
fully agreeable INDO
results for both HQ$_{pcff}$-HQ$_{cfvv}$ and
 HQ$_{cfvv}$-HQ$_{cfvv}$ sets of data. 
To check for  localization of the HOMO and LUMO orbitals a simple routine has been
undertaken:
In the LCAO framework each molecular orbital ($\psi_i$) is composed as a linear 
combination of atomic orbitals ($\phi_k$) $\psi_i = \sum_k c_{ik} \phi_k$.
The ratio of the sum of squared coefficients $c_{ik}^2$ over atomic orbitals of one 
of the monomers  to the relevant sum for all atoms in a dimer
has been assumed to be a measure of the localization of the $i-th$ molecular 
orbital:
\begin{equation}
\mbox{orbital localization} =\frac{\sum_{\mbox{\scriptsize \it one monomer}} c^2_{ik}}{\sum_{\mbox{\scriptsize \it both monomers}} c^2_{ik}}.
\end{equation}
Based on the density matrix, all charges and their distributions have been calculated
according to the Mulliken population analysis.
\section{Results}
\subsection{Neutral dimers}
\begin{figure}
\begin{center}
\epsfig{file=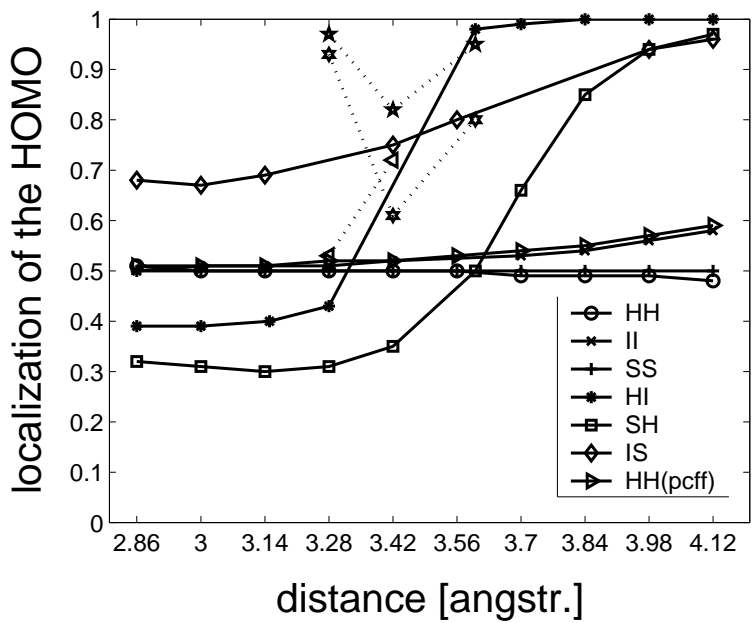,width=10cm}
\caption{Localization of the HOMO orbital as defined in eq.(1) for the neutral 
aggregates. Data are displayed for the
HQ monomer in the case of Hi and SH assemblies and for the SQ monomer in the
case of the SI dimer. Dotted lines connect extra data points for
the SH dimer with altered configurations: $\lhd$ and heksagrams - the top SQ
 molecule rotated by
90$^\circ$ and 180$^\circ$, respectively; pentagrams - the top molecule shifted
along $x$-axis, so that the 6-membered ring of the molecule was overlapping the
5-membered ring of the another monomer.}
\label{homos}
\end{center}
\end{figure}

\begin{figure}
\begin{center}
\epsfig{file=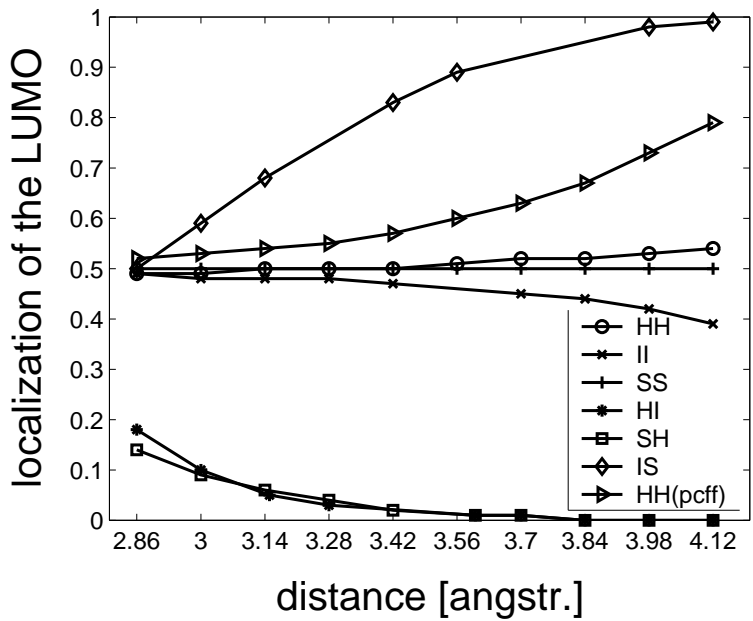,width=10cm}
\caption{Localization of the LUMO orbital on the neutral charged agregates.
Data points displayed for the HQ monomer in the case of HI, SH dimers and for
the SQ monomer in the case of SI dimers.}
\label{lumos}
\end{center}
\end{figure}

In the absence of any extra charge both monomers remain neutral for all
 distances reported in
this study. The 
biggest value for charge separation has been observed for SQ-HQ set at the 
distance\footnote{Intermolecular distances have been defined as distances between the 
mean planes
located on all atoms of the separated monomers.} 
2.86 \AA$ $ (charge 0.26 and -0.26 respectively).
The degree of orbital localization as defined above  is displayed in Figuress~\ref{homos} 
and \ref{lumos}.

The results of the analysis are displayed as HOMO (LUMO) localization for
 HQ monomer in the
 sets  of HS, IH dimers and  as HOMO (LUMO) localization for SQ monomer in
the IS dimers, respectively.
By inspection of Figure~\ref{homos} one can deduce that in the case of symmetrical 
assemblies HOMO is delocalized even at relatively large separations of monomers.
This tendency remains also unchanged for HQ$_{cvff}$-HQ$_{pcff}$
sets, for which the spatial symmetry of coordinates is broken.
 At extremal intermolecular distances ($\approx$ 10 \AA) the HOMO
orbital becomes localized on  one of the monomers that has been examined 
for both 
HQ$_{cvff}$-HQ$_{pcff}$ and HQ$_{cvff}$-HQ$_{cvff}$ dimers sets.
However, the delocalization decreases faster with encreasing distance 
when the mixed
sets of dimers (SH, HI, IS) are analyzed. Distances of about 3.28-3.7\AA{} 
seem to be "transition distances": If the monomers are closer to each other,
they behave as supermolecule, if they are farther apart, they become two separated
molecules.\\ 
Additional examination provides further arguments that at large distances
 ($\approx 4$\AA),
the supermolecular structure of dimers breaks down and mono\-mers behave
as two weakly interacting molecules. By analyzing the five largest coefficients $c_{ik}$ 
and the
energy of the relevant orbitals we have found that at the distance 4.12 \AA{} 
HOMO-1/HOMO/LUMO/LUMO+1 aggregates' orbitals are created of HOMO/LUMO relevant monomers'
orbitals. (see Tab.~\ref{tabelka}).
 
\begin{table}
\begin{center}
\begin{math}
\begin{array}{l|l|l}
\hline
\hline
HI&SH&IS\\
\hline
HI_{HOMO-1} \approx IQ_{HOMO}&SH_{HOMO-1} \approx SQ_{HOMO}&IS_{HOMO-1} \approx IQ_{HOMO}\\
HI_{HOMO} \approx HQ_{HOMO}&SH_{HOMO} \approx HQ_{HOMO}&IS_{HOMO} \approx SQ_{HOMO}\\
HI_{LUMO} \approx IQ_{LUMO}&SH_{LUMO} \approx SQ_{LUMO}&IS_{LUMO} \approx SQ_{LUMO}\\
HI_{LUMO+1} \approx HQ_{LUMO}&SH_{LUMO+1} \approx HQ_{LUMO}&IS_{LUMO+1} \approx IQ_{LUMO}\\
\hline
\hline
\end{array}
\end{math}
\vskip 0.5truecm
\caption{Structure of the two highest occupied and two lowest unoccupied
molecular orbitals of the indolic sandwich at 4.12\AA{} intermolecular
separation.}
\label{tabelka}
\end{center}
\end{table}

When the intermolecular distance decreases, HOMO of the aggregate
becomes a mixture of the HOMOs of both  monomers.
At the same time the HOMO-LUMO gap does not change and remains within the range 0.20-0.22a.u.
 for HI and
0.19-0.20a.u.  for SH dimers, respectively. However, in the case of the IS
  dimers
a monotonic decrease in the energy of the HOMO-LUMO gap is registered (from 0.20a.u.
 at 4.12 \AA{}
to 0.16 at 2.86 \AA).
Figure~\ref{homos} comprises also an extra data set for SH dimers:
In one case the molecules have been shifted apart preserving the same intermolecular 
distance - a bigger ring of one molecule was above the smaller
ring of the other. In the second case, one of the molecules has been
 rotated in plane by 90$^{\circ}$ or 180$^{\circ}$ (monomers remained parallel). 
The results indicate that such a variation in the
configuration of a dimer  changes significantly localization of orbitals,
although none persistent tendencies have been observed.
In contrast to HOMO, LUMO orbitals reveal lower tendency to delocalization.
 Only LUMO of the IS aggregate 
is more delocalized (at small intermolecular distances) than its HOMO.
\subsection{Charged dimers}
\begin{figure}
\begin{center}
\epsfig{file=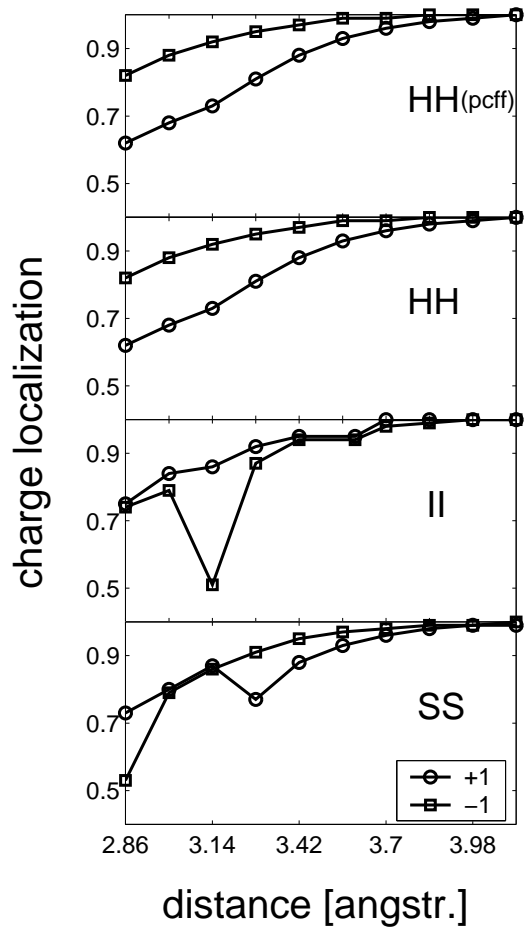,width=7cm}
\caption{Localization of the additional charge on the agregates ($\circ$ for negative charged,
$\Box$ for positive charge, $\lhd$ for negative charged HQ$_{cvff}$-HQ$_{pcff}$,
$\rhd$ for positive charged HQ$_{cvff}$-HQ$_{pcff}$ assemblies).}
\label{charges}
\end{center}
\end{figure}

\begin{figure}
\begin{center}
\epsfig{file=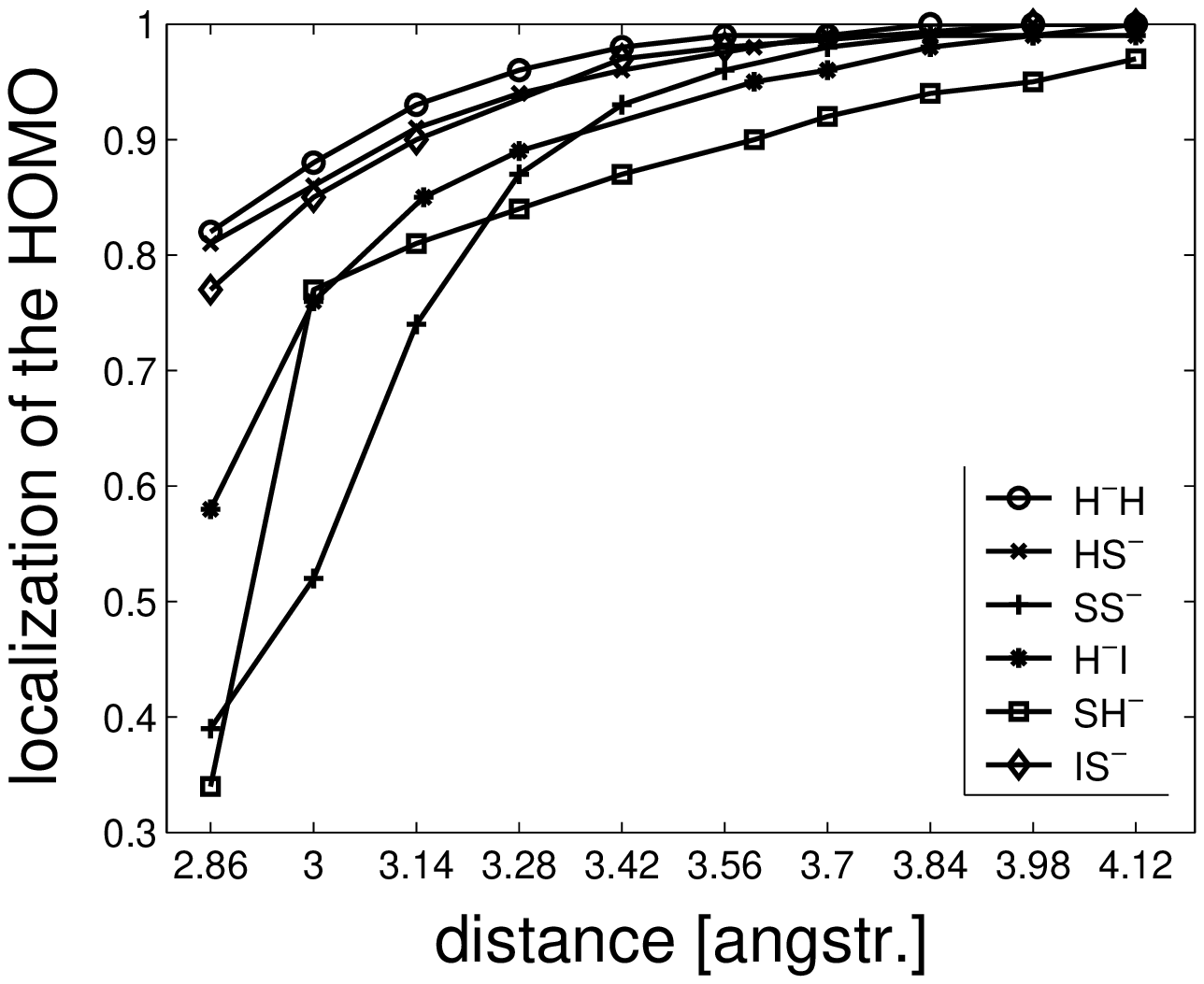,width=10cm}
\caption{Localization of the HOMO for assemblies with one proton removed from
the
 -OH group
of one of the monomers. Data points are displayed for the monomer without the proton.}
\label{homo_b}
\end{center}
\end{figure}

\begin{figure}
\begin{center}
\epsfig{file=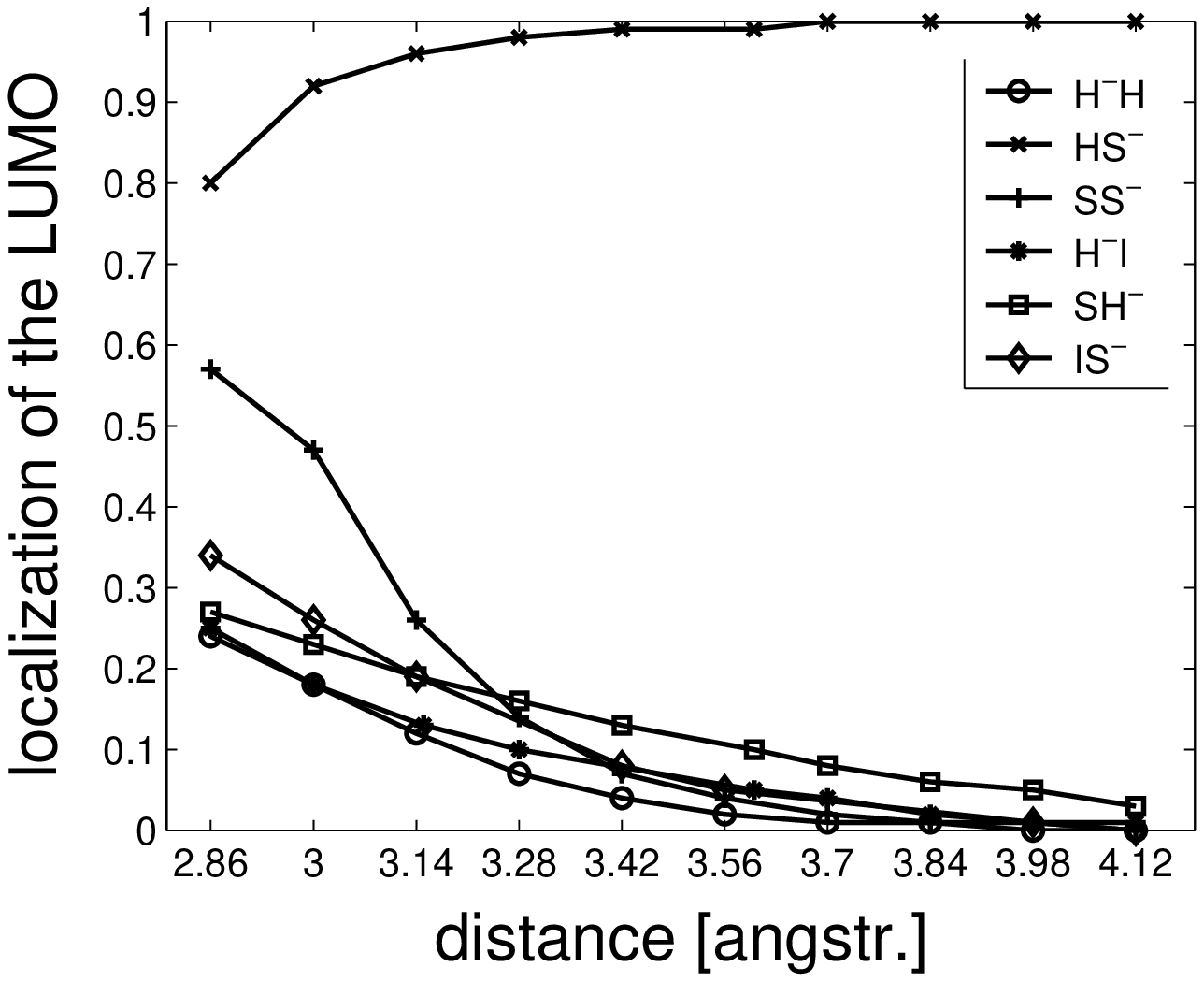,width=10cm}
\caption{Localization of the LUMO for assemblies with one proton removed from
the -OH group
of one of the monomers. Data points are displayed for the monomer without the proton.}
\label{lumo_b}
\end{center}
\end{figure}

In case of molecules with unchanged number of atoms both, negatively and positively 
charged assemblies have been examined. 
Charge localization for symmetrical 
sets (HH, II, SS) are displayed in the Figure~\ref{charges}. 
As can be inferred from the latter, an extra charge placed on a dimer is localized
on one of the monomers, although as expected,  the smaller the distance, 
the tendency 
of charge delocalization  becomes more noticeable.\\ 
For negatively charged heterogeneous assemblies (SH, HI, IS), at all 
intermolecular distances the charge is localized on the more electron-liked monomer, 
{\it i.e.} (SH)$^-$ results in localization of charge on S monomer, (SH)$^-$ = S$^-$H
and similarily, (HI)$^-$ = H$^-$I and (IS)$^-$ = IS$^-$.\\ 
This finding changes however when a positive
charge is taken into account. The molecules are poor electron-donors and
our results show  that the positive charge becomes localized on  one of the monomers without any
correlation with distance or monomer's redox state. \\
For negatively charged systems the localization of the HOMO orbital has been also examined
indicating that the orbitals are restrained exactly to the same monomer as 
the charge itself.\\
Similar behavior has been observed for dimers of molecules with a proton removed
from the -OH group of one of the constituing monomers ({\it cf.}
Figures~\ref{homo_b}, \ref{lumo_b}). 

\begin{figure}
\begin{center}
\epsfig{file=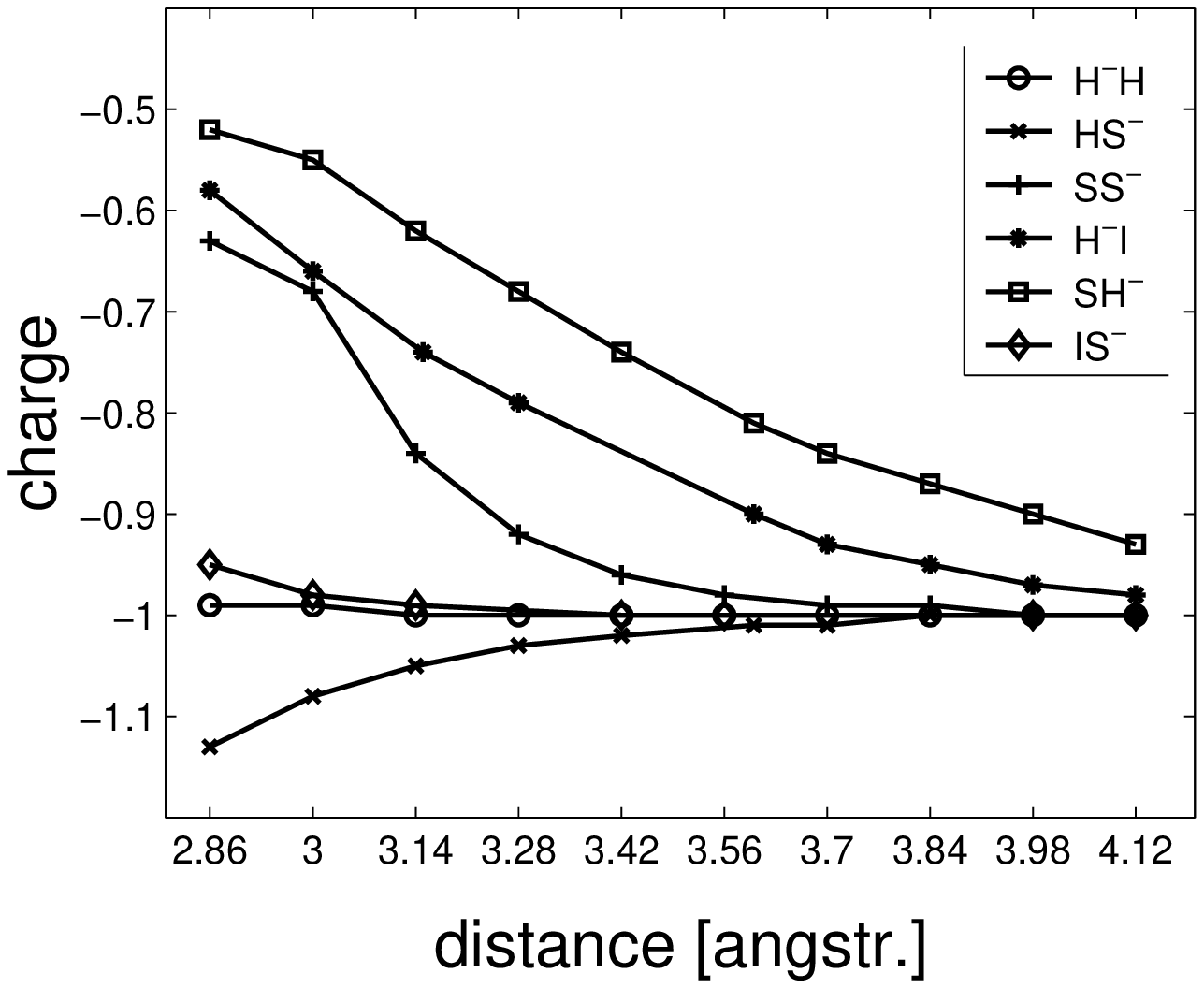,width=10cm}
\caption{Charge localization for assemblies with one proton removed from -OH group of one
of the monomers. Data for the monomer without the proton.}
\label{charges_b}
\end{center}
\end{figure}

With the  exception of the HS-dimer sets, all HOMOs for deprotonated assemblies
are localized on different monomer than the relevant LUMOs 
({\it cf.} Figures ~\ref{homo_b}, \ref{lumo_b}). At the same time negative charge is 
localized on the same monomer as the relevant HOMO of the dimer
(Figure~\ref{charges_b}). 
The results of our computational analysis point out a persistent tendency
for localization of charge/HOMO/LUMO within the studied dimeric assemblies of DHI 
monomers at the separation exceeding 3.4\AA{}.
 Following the observation of  HOMOs and LUMOs localization in the dimer, we
 have also examined
 the structure of first excited states of the assemblies. In most cases, the
 first excited state of a dimer  is composed of its relevant 
HOMO$\rightarrow$LUMO excitation, 
the apparent  exceptions are however HI dimers at 3.00--3.14 \AA{} separation
 and 
SH dimers at 3.28\AA{} separation whose first excited states are mostly composed
of the HOMO-1$\rightarrow$LUMO transitions.
Similarily, the first excited state of the SH dimer at 3.00--3.14\AA{}
 separation and of the IS dimer at 3.42\AA{} (and more) separation are 
 predominantly formed of the HOMO-1$\rightarrow$LUMO and HOMO$\rightarrow$LUMO+1
  transitions, respectively. 
\section{Discussion and Summary}
The simplest sets of the stacked monomers, building blocks of eumelanin polymers have 
been examined. Analysis of localization of additional charges and 
HOMO/LUMO orbitals in an indolic  dimer structure has been performed.
Charged aggregates have been created either by adding/removing one electron
from the neutral dimeric structure,  or by
deprotonating the -OH group in one of the monomers. Our investigations
 show that the structure of the frontier orbitals  of any 
particular dimer  depends on redox state of its 
components as well as on the way the molecule has gained the charge. Heterogeneous structures
(SH, HI, IS) tend to  behave as two separate molecules 
even if the distance between the molecules is small. 
Moreover, negatively charged dimers formed by adding an excessive electron
to the neutral aggregate 
are characterized by the charge localization on a more electron-like monomer.
In dimers where one of constituing monomers has been deprotonated,  
the proton-deficient monomer  holds the negative charge 
independent of the electron-affinity of the neighbouring molecule in the set.\\
 Homogeneous sets (HH, II, SS) of monomers show a
stronger  tendency to create dimers than the respective heterogeneous assemblies
and at
 comparable distances they behave as one
supermolecule with its HOMO and LUMO orbitals smeared over the whole structure. 
Additional charge in this structures becomes, however,  located on one monomer
 only. Similar conclusion can be drawn from the preliminary studies on sets of
 planar pentameric polymers  formed of indolic units, whose structure and
 absorption properties have been discussed elsewhere \cite{my}.
 Same analysis performed for neutral and charged assemblies of that type 
 has revealed that also in the pentamers of randomly assemblied
 HQ,IQ and SQ molecules, the excessive charge becomes localized on one of 
 constituing monomers.   
This result suggests that, in spite of the electron-acceptor properties, 
the indolic 
dimers (and most likely, indolequinone polymers, in general) do not stabilize
 the charge by delocalization over the entire structure, but rather tend to
position the charge on one of the composing molecules.\\
 The additional analysis performed on the SH dimer sets with 
one molecule rotated or translated with respect to another suggests that
 electronic 
properties of the complex may strongly depend on its conformations.
Nevertheless, if we limit our considerations to characteristic distances (3.4\AA) 
reported in experimental studies, the aformentioned general conculsions 
remain valid.\\ \\

The above findings along with the conclusions recalled from References
\cite{galvao1,galvao2,prota2,sarna} suggest a highly dynamic picture of
hypothetical eumelanins, whose chemical structure has been proven to be composed
mainly of 
DHI molecules in different redox states. The redox state of the melanin units
may change spontaneously under the influence of the environment (acidity, exposure to
light and oxygen) and their relevant proportion and state of oxidation
may vary in time.
In agreement with the former theoretical studies \cite{marinez}, our results 
show high charge localization in basic clusters of eumelanin structure.
Given the latter and an apparent localization of the HOMO and LUMO orbitals 
on one of the monomers in the stack, the electron-acceptor behavior of the melanin can be 
viewed as a process in which an electron injected (or photogenerated) in
the polymer remains trapped or transfers {\it via} a hopping mechanism
in between the indolic subunits.
Moreover, any temporary charge distribution within the eumelanin complex
can be altered {\it e.g.}  in the process of spontaneous HQ$\rightarrow$IQ reaction
that effectively will also influence structure of frontier orbitals of
 the aggregate. Discussed from that perspective, eumelanins can be understood as
 amorphous polymers whose macroscopic electronic properties are essentially ensemble-average
  of various  structures.
  \section{Acknowledgements}
  The authors acknowledge the Research Grant (1999-2002) from the
  Marian Smoluchowski Institute of Physics, Jagellonian University.


\end{document}